\documentclass[aps,prl,twocolumn,superscriptaddress]{revtex4-1}

\usepackage{fancyhdr} 

\usepackage{hyperref} 

\usepackage{verbatim} 
\usepackage{amsmath} 
\usepackage{amsthm} 
\usepackage{amssymb}
\usepackage{graphicx} 
\usepackage[dvips,letterpaper,margin=0.75in,bottom=0.5in]{geometry}

\let\baraccent=\= 
\renewcommand{\=}[1]{\stackrel{#1}{=}} 

\newcommand{\gae}{\lower 2pt \hbox{$\,
    \buildrel{\scriptstyle >}\over {\scriptstyle \sim}\,$}}
\newcommand{\lae}{\lower 2pt \hbox{$\,
    \buildrel{\scriptstyle <}\over {\scriptstyle \sim}\,$}}

\newcommand{\abs}[1]{\left| #1 \right|} 
\newcommand{\avg}[1]{\left< #1 \right>} 
 
\newcommand{\ket}[1]{\big| #1 \big\rangle} 

\usepackage{color}
\usepackage{ulem}

\begin{document}

\title{Universal Dynamics of Stochastically Driven Quantum Impurities}

\author{William Berdanier}
\email[]{wberdanier@berkeley.edu}
\affiliation{Department of Physics, University of California, Berkeley, CA 94720, USA}
\affiliation{Kavli Institute for Theoretical Physics, University of California, Santa Barbara, CA 93106-4030, USA}

\author{Jamir Marino}
\affiliation{Department of Physics, Harvard University, Cambridge, MA 02138, USA}
\affiliation{Kavli Institute for Theoretical Physics, University of California, Santa Barbara, CA 93106-4030, USA}
\affiliation{Department of Quantum Matter Physics, University of Geneva, 1211, Geneva, Switzerland}

\author{Ehud Altman}
\affiliation{Department of Physics, University of California, Berkeley, CA 94720, USA}
\affiliation{Kavli Institute for Theoretical Physics, University of California, Santa Barbara, CA 93106-4030, USA}

\date{\today}

\begin{abstract}
We show that the dynamics of a quantum impurity subject to a stochastic drive on one side and coupled to a quantum critical system on the other display a universal behavior inherited from the quantum critical scaling. Using boundary conformal field theory, we formulate a generic ansatz for the dynamical scaling form of the typical Loschmidt echo and corroborate it with exact numerical calculations in the case of a spin impurity driven by shot noise in a quantum  Ising chain. We find that due to rare events the dynamics of the mean echo can follow very different dynamical scaling than the typical echo for certain classes of drives. Our results  are insensitive to irrelevant perturbations of the bulk critical model and apply to all the microscopic models in the same universality class.
\end{abstract}
\maketitle

Universality lies at the core of our understanding of equilibrium critical phenomena and is successfully captured by the renormalization group framework~\cite{Cardy:1996xt,sachdev2007quantum}. 
This program has been extended to non-equilibrium classical systems, leading to the discovery of new dynamical universality classes, including coarsening, reaction-diffusion, and surface growth, among several others~\cite{Tauberbook2014}. Recent developments in experiments with quantum many-body systems call for a further extension of the program to universal phenomena in quantum dynamics. For example, systems of ultracold atoms and ions exhibit new dynamical transitions~\cite{PhysRevLett.119.080501, 1806.11044, zhang}, as well as new forms of dynamical scaling~\cite{PhysRevLett.115.245301, ober, schmied, cor}. Other classes of universal phenomena are seen in driven open quantum systems. These include experiments with  non-equilibrium Bose-Einstein condensation of polaritons~\cite{Kasp2006}, dissipative phase transitions  in cavity QED circuits~\cite{PhysRevX.7.011016}, and dynamical phase diagrams of condensates trapped in optical cavities~\cite{PhysRevA.99.053605, klinder2015dynamical}.
The common wisdom is that  driven-dissipative quantum  systems exhibit emergent classical dynamics  because the coupling to the environment  washes out the delicate quantum coherences. For instance, the occurrence of effective Langevin dynamics is common to many quantum systems coupled to a bath, with examples ranging from cold atoms to
solid state platforms~\cite{sieb, mitra06, PhysRevB.85.184302}. In certain cases an intermediate regime of universal quantum scaling can be identified~\cite{PhysRevB.85.184302,PhysRevB.94.085150}, but it remains an open question whether such quantum scaling can persist to all scales in a driven-dissipative system.

In this Letter,  we show that universal, inherently quantum scaling can emerge in a conformally invariant system driven out of equilibrium by a stochastic boundary field.  We consider microscopic models with Hamiltonian of the form

\begin{equation}\label{cftham}
\hat{H}=\hat{H}_{\mathrm{CFT}}+{h}_b(t)\hat{O}_b,
\end{equation}

where $\hat{H}_\mathrm{CFT}$ is a one-dimensional bulk  critical Hamiltonian driven by a stochastic noise field $h_b(t)$, weighted by a relevant operator $\hat{O}_b$ that lives on the boundary of the system. 
Generically, $\hat{H}_\mathrm{CFT}$ can include irrelevant terms that break the conformal symmetry, and only emergent conformal invariance in the infrared limit is required. 
Previous work investigated the coupling of quantum systems to different types of boundary drives, which lead to eventual thermalization~\cite{prosen,PhysRevB.85.184302} or to non-universal relaxation~\cite{PhysRevLett.122.040604}; in contrast, we show that the dynamics induced by a conformal boundary drive are universal in a certain limit and inherently quantum. 

Before proceeding, we note that the problem of a CFT driven by a periodic (Floquet) boundary drive, considered by one of us \cite{PhysRevLett.118.260602},  does lead to universal relaxation. In this work we find that universality persists even with a more generic stochastic drive. Furthermore, we show that the behavior of the Loschmidt echo is richer than in the periodically driven case: one may have a different class of universal relaxation when looking at the typical decay in a single realization of the noise compared to the average echo over many noise realizations. 
We corroborate these results with a direct numerical calculation of a boundary driven transverse field Ising model at its critical point.

\begin{figure}
    \centering
    \includegraphics[width=\columnwidth]{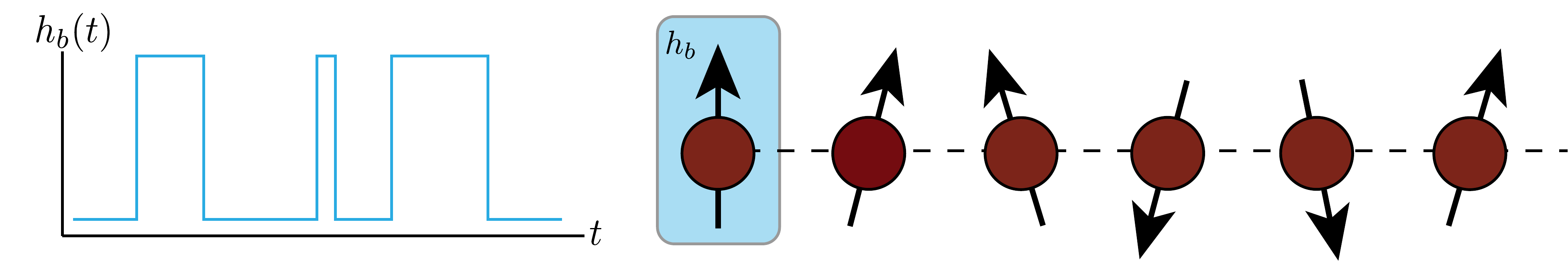}
    \caption{Sketch of the class of systems under study in this work. We consider a quantum critical spin chain (red spins) subjected to a stochastic boundary drive (blue line).}
    \label{fig:model}
\end{figure}

\paragraph{Stochastically driven boundary in CFT.} For concreteness, let us consider the Poisson process whereby the boundary coupling $h_b$ stochastically jumps between two values with some fixed probability $p$ over an interval of time $\delta t$ as illustrated schematically in Fig. \ref{fig:model}. We note that generically any type of sufficiently weak Markovian noise will flow to Poisson noise under the renormalization group (RG), since events that flip the boundary conditions are more relevant than those that do not. To define our scaling variables, we have an average time between flips $T = \delta t/p$, with a Poisson parameter of the shot noise after total time $t$ of $\lambda=pt/\delta t$. Finally, the strength of the boundary field  $h_b \sim ||h_b(t)||$ sets the timescale $t_b = h_b^{-\nu_b}$. Here, $\nu_b = 1/(1-\Delta_b)$ with $\Delta_b$ the scaling dimension of the boundary operator $\mathcal O_b$. Application of the boundary CFT framework is valid while the time between flips of boundary conditions is much larger than the timescale $t_b$, hence the latter serves as a short time cutoff for our theory.

In what follows we focus on the Loschmidt echo, or return-probability amplitude of the wavefunction,
\begin{equation}\label{eq:los}
\mathcal{L}(t)=|\langle \psi_0|\psi(t)\rangle|^2,
\end{equation}
{which in recent years has become an important quantity for the study of universal properties of quantum many-body systems~\cite{dpt, PhysRevLett.119.080501,PhysRevB.99.134301,Silva08} and can be measured via spectroscopic techniques~\cite{adilet, Latta11}.}
We consider the behavior of this function in a typical realization of the stochastic drive field as well as its expected average over all possible realizations of the noise. 

For each realization of the stochastic field $h_b(t)$, $\mathcal{L}(t)$ can be mapped to a partition function of a field theory, {which flows to a conformally invariant one in the scaling limit where the time between flips is much larger than $t_b$.}~\footnote{This generalizes the constructions of Ref.~\cite{PhysRevX.4.041007} and Ref.~\cite{PhysRevLett.118.260602}.}. After a Wick rotation to imaginary time, the ground state $\ket{\psi_0}$ is determined as the asymptotic evolution $\lim_{\tau \to \infty} e^{-\tau H_0} \ket{\Omega}$,  with $\ket{\Omega}$ a generic state and the operator $e^{-\tau H_0}$ acting as a projector onto the ground state of $H_0$ in the limit $\tau\to\infty$. The boundary field flips between different fixed values at random times; therefore, in any given realization of the flips, the unitary time evolution operator takes the form of a succession of imaginary time evolutions, given by the Hamiltonian~\eqref{cftham} with different fixed boundary fields over the intervals between flips. Thus, we have $\mathcal{L}(t) \propto |\avg{e^{-\tau_0 H_0} \ldots e^{-\tau_2 H_2} e^{-\tau_1 H_1} e^{-\tau_0 H_0}}|^2$, with $\tau_0\to\infty$. Since the Hamiltonians $H_i$ differ only by a relevant boundary operator, we see that this maps exactly onto a partition function in a two-dimensional conformal field theory with mixed boundary conditions along the imaginary time direction. 

Now let us focus on the case of $T \gg t_b$, that is, the average time between flips being much greater than the timescale induced by the finite boundary field. This is to ensure that the dynamics enters into a universal regime where it can exhibit scaling.  It is also important that we impose $\delta t \gtrsim t_b$, since we only expect universal physics on timescales longer than $t_b$, and $\delta t$ is the minimal spacing between flips. These limits allow us to use the technique of boundary condition changing operators, generic to any two dimensional conformal field theory, in which sharp changes in the boundary condition may be replaced inside all correlation functions by a particular type of primary operator, often referred to as a boundary-condition changing (BCC) operator, inserted at the location of the change~\cite{cardy_conformal_1984, cardy_boundary_1989, cardy_boundary_2005}. We can therefore identify the Loschmidt echo with a $2n$-point function of primary operators $\phi_{\mathrm{BCC}}$. Analytically continuing to real time, for any realization of the noise with flips at times within some configuration $S=\{t_i \}$, the Loschmidt echo is then

\begin{equation}
\mathcal{L}(t|\{t_i\}) \sim \Big| \Big\langle \prod_{t_i \in S}\phi_{\mathrm{BCC}}^{(i)}(t_i) \Big\rangle \Big|^2.
\end{equation}

For simplicity, let us now assume that we have a binary drive between two Hamiltonians $H_0$ and $H_1$, and hence only one type of BCC operator, $\phi$, per drive, though we note that the argument follows for more complicated drives as well. The specific examples that we consider below are boundary drives in the critical Ising model. One class of drive in this case is given by a boundary condition that jumps back and forth between fixing the boundary spin up/down. We call this the ``fixed-fixed'' drive, and it corresponds to insertions of a fermion BCC operator with scaling dimension $\Delta_{\mathrm{BCC}}=1/2$. Another class of drive is given by a field that jumps between a free and fixed (say, spin up) boundary condition. This drive corresponds to inserting a BCC operator with a scaling dimension $\Delta_{\mathrm{BCC}}=1/16$.

{\it The typical echo.} We first calculate the typical echo $\mathcal{L}_{\mathrm{typ}} \equiv e^{\overline{\log \mathcal L}}$. We have

\begin{equation}\label{eq:typ}
\overline{\log \mathcal L} = \sum_{n=0}^\infty P(n) \frac{1}{t^n/n!} \int \prod_i dt_i \log \abs{\mathcal{C}(t_1, ..., t_n) }^2,
\end{equation}

with $\mathcal{C}(t_1, ..., t_n)=\avg{\phi(t_n) \ldots \phi(t_1)}$ the time-ordered correlation function associated to $n$ insertions of the BCC operators, $P(n) = e^{-\lambda} \lambda^n / n!$ for a Poisson process, and we note that in Eq.~\eqref{eq:typ} only $2n$-point functions enter the expectation value. In fact, because of the ket in the echo, both the one-flip process and the two-flip process are controlled by the two-point function of BCCs, and similarly for higher orders: the $(2n-1)$- and $2n$-flip processes are controlled by the $2n$-point function of BCCs. In taking the average over the BCC insertions, we normalize by $\int \prod_i dt_i = t^n/n!$, where the $n!$ factor is due to the time-ordering. 

Now, for average flipping times $T$ much larger than the microscopic timescale $t_b$ ($T\gg t_b$), we can utilize the finite-size scaling relation for primary operators at the bulk critical point~\cite{Cardy:1996xt}, i.e. $|\mathcal{C}(t_1, ..., t_n)|^2 =  (T/t_b)^{-4n\Delta_{\mathrm{BCC}}} \mathcal F(t_1/T,\ldots, t_n/T)$, with $\mathcal F$ a universal scaling function. We therefore expect the typical Loschmidt echo to be a universal scaling function $\mathcal L_{\mathrm{typ}} = \mathcal L_{\mathrm{typ}}(T/t_b, \lambda)$, and after explicit evaluation of the sum we arrive at

\begin{figure*}
	\includegraphics[width=\columnwidth]{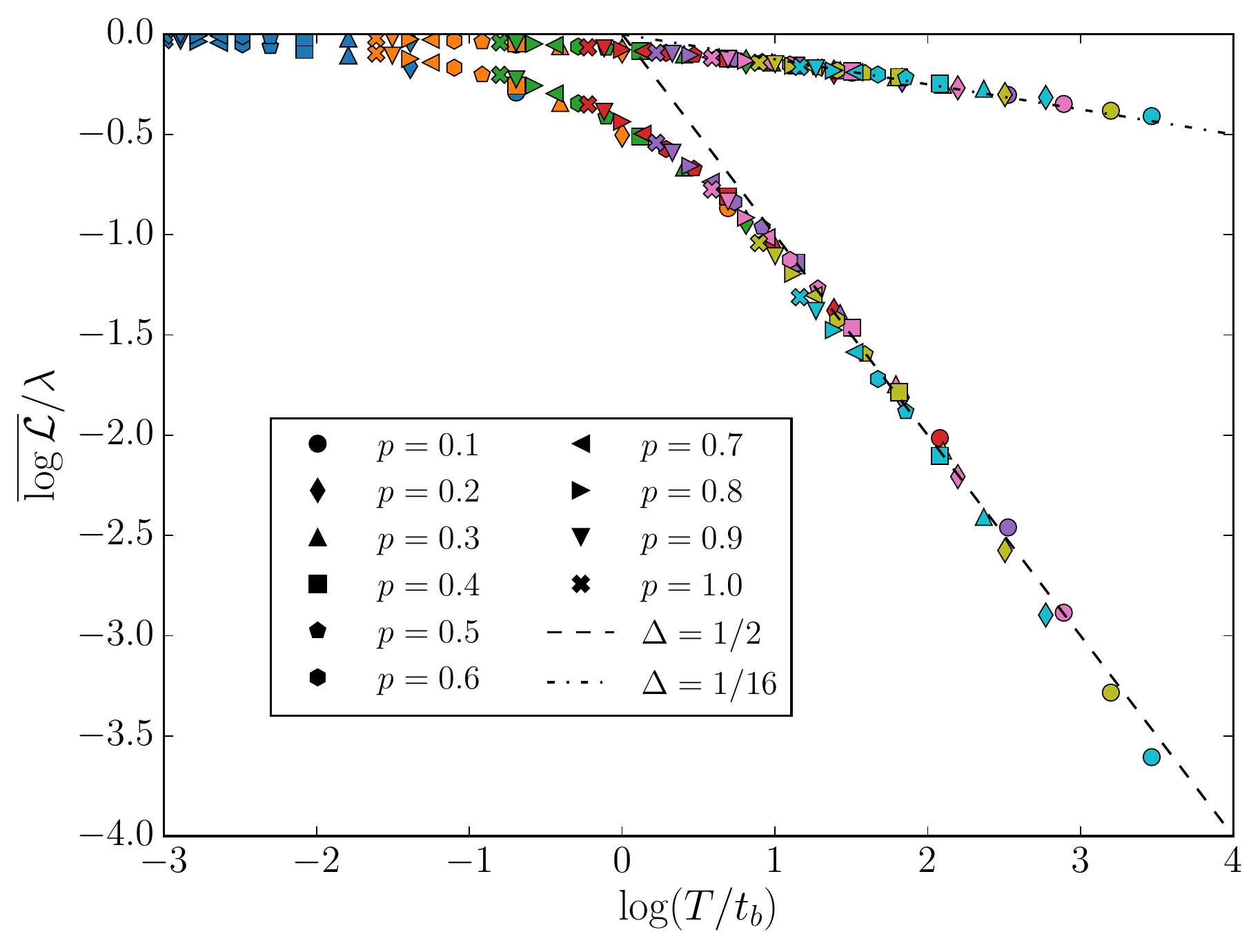}
	\includegraphics[width=\columnwidth]{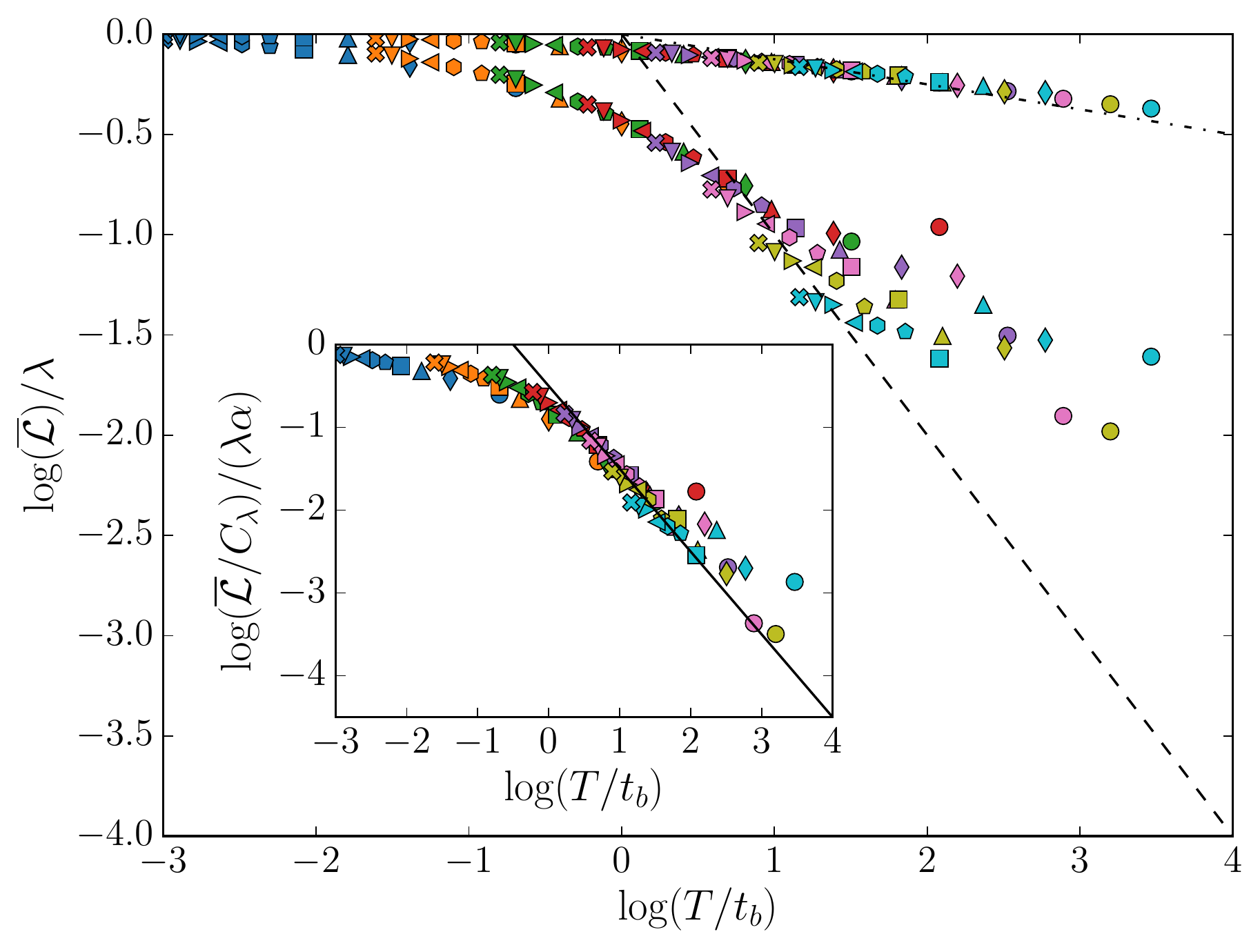}
	\caption{\label{fig:data} Left: The typical Loschmidt echo averaged over $r=1000$ realizations and for  system sizes up to $L=1000$, for different values of the boundary field $h_b$ and flipping probabilities $p$. The boundary field takes the values $h_b = 0.1$ (blue), 0.2 (orange), 0.3 (green), 0.4 (red), 0.5 (purple), 0.6 (pink), 0.7 (yellow), 0.8 (teal), and the probability $p$ varies as marked in the legend. The Poisson parameter $\lambda$ takes values $\lambda \geq 10$ throughout. The dashed lines are the prediction from boundary CFT (Eq.~\eqref{scaling}): for the fixed-fixed drive $\Delta = 1/2$, and for the free-fixed drive $\Delta = 1/16$, with both showing excellent agreement. Right: The mean echo of the same data. For the free-fixed drive, the mean and typical (black dashed lines) are very similar, but, strikingly, for the fixed-fixed drive the mean lies far above the typical. This is due to rare events that dominate the average and give a renormalized scaling form (inset), where $\alpha = 0.71\pm 0.03$, in good agreement with the estimate of $\alpha = 0.75$ in the main text.  
	}
	\label{fig:numerics}
\end{figure*}

\begin{widetext}
\begin{equation}\label{eq:scaling2}
\overline{\log \mathcal L} \simeq -4 \Delta_{\mathrm{BCC}} \sum_{n=1}^\infty (P(2n-1) + P(2n))n \log(T/t_b)= -2\Delta_{\mathrm{BCC}} (\lambda + e^{-\lambda} \sinh \lambda) \log(T/t_b),
\end{equation}
\end{widetext}

up to an additive universal average amplitude term $\overline{\log \mathcal F}$ that may be neglected in the large $T$ limit. We note that averaging the logarithm is crucial, as the amplitude itself may in general diverge. For large $\lambda \gg 1$ we expand this result to obtain $\overline{\log L} \approx_{\lambda \gg 1} -2\Delta_{\mathrm{BCC}} \lambda  \log(T/t_b)$. Thus, we predict a universal power-law form of the typical echo

\begin{equation}\label{scaling}
\mathcal{L}_{\mathrm{typ}}  \underset{\lambda\gg 1}{\sim} \left(\frac{T}{t_b}\right)^{-2\Delta_{\mathrm{BCC}} \lambda},
\end{equation}
which is in good agreement with the numerical data on the Ising model shown in Fig.~\ref{fig:data}.

\paragraph{Mean echo.} Having argued for universal behavior of the Loschmidt echo in a typical realization of the boundary stochastic field, we now turn to the calculation of the mean echo. 
In many cases, the mean echo should follow the same universal scaling form as the typical. However, as we argue below, for certain types of drives the mean and typical echo may differ drastically.

The general form of the mean echo is given by 

\begin{equation}
\overline{\mathcal L(t)} = \sum_n P(n) \frac{1}{t^n/n!}\int \prod_{k=1}^n dt_k \abs{\avg{\phi(t_n) \ldots \phi(t_1)}}^2 .
\label{eq:avL}
\end{equation}

As noted previously, finite-size scaling implies that for $T\gg t_b$ the $n$-point function should be a power law in $T$, with an exponent determined by the scaling dimension of the BCC operator.
If $\Delta_{\mathrm{BCC}} \geq 1/4$, the power-law can produce a divergence in (\ref{eq:avL}) when integrating over the insertions of the BCCs. In the divergent case, rare configurations where the insertions are all closely spaced can give a dominant large contribution to the mean echo, while they do not affect the typical echo because the integral is over the logarithm. 

Let us show this explicitly. Consider first the case $\Delta_{\mathrm{BCC}} < 1/4$, where the integrals are non divergent. An example is the fixed-free drive of the Ising model with $\Delta_{\mathrm{BCC}}=1/16$.
Using the aforementioned finite-size scaling relation $|\mathcal{C}(t_1, ..., t_n)|^2 =  (T/t_b)^{-4n\Delta_{\mathrm{BCC}}} \mathcal F(t_1/T,\ldots, t_n/T)$, we have $\overline{\mathcal L(t)} = \sum_n P(n) (T/t_b)^{-4n \Delta_{\mathrm{BCC}}} \overline{\mathcal F(n)}$, where $\overline{\mathcal F(n)} = \int \prod dt_i \mathcal F(t_1/T,\ldots, t_n/T)$ is finite and independent of the lower cutoff. This sum can be evaluated using the saddle point approximation. One finds that, under the assumption $\lambda \gg T/t_b$, the sum is dominated simply by the term $n_* = \lambda / 2$, recalling that the sum runs only over $n$ even. Therefore, one obtains $\overline{\mathcal L(t)} \sim e^{-\lambda/2} \overline{\mathcal F(\lambda/2)}  (T/t_b)^{-2\Delta_{\mathrm{BCC}} \lambda}$. This gives the same power law dependence on $T$ as the typical echo, and hence the same scaling form. 

Now consider the divergent case $\Delta_{\mathrm{BCC}}\ge 1/4$, which is realized, for example, by the fixed-fixed drive of the Ising model ($\Delta_{\mathrm{BCC}}=1/2$). 
In this case the integral over $\mathcal F(t_1/T,\ldots, t_n/T)$ depends sensitively on the lower cutoff $\delta t/T$. In order to estimate of the scaling form, we replace the averaged correlation function of BCCs by the largest contribution in the limit $\delta t \to 0$. Namely, we take $\int \prod dx_k {\mathcal F}(x_1, ..., x_n) \approx (\delta t/T)^{(1-4\Delta_{\mathrm{BCC}})n/2}$, where $\delta t^{1-4\Delta_{\mathrm{BCC}}}$ is the divergent part of the two-point function. Substituting this into the sum and taking $\delta t \approx t_b$ gives $\overline{\mathcal L} \sim \sum_n P(n) \frac{n!}{\lambda^n} (T/t_b)^{-4n\Delta_{\mathrm{BCC}}} (t_b/T)^{{n\over 2}(1-4\Delta_{\mathrm{BCC}})}$. Finally, using the saddle point method with the sum dominated by $n_* = \lambda / 2$, we obtain the power law $\overline{\mathcal L} \sim (T/t_b)^{-\lambda(\Delta+{1\over 4})}$, which is different than power law governing the typical echo. In particular, for the fixed-fixed Ising drive we get $\overline{\mathcal L}\sim (T/t_b)^{-3\lambda/4}$, which should be compared with ${\mathcal L}_{\mathrm{typ}}\sim (T/t_b)^{-\lambda}$.

\paragraph{Numerical results.} Having expounded our arguments in generality for stochastically boundary-driven CFTs, let us now validate them in an explicit model. Consider a one dimensional integrable quantum Ising chain in a transverse field, $g$, tuned to criticality, $g\to g_c$, and driven by a stochastic time-dependent noise coupled to the longitudinal spin field at the boundary of the chain, 

\begin{equation}\label{eq:ham}
H(t) = - J\sum^L_{i=1} ( \sigma_i^z \sigma_{i+1}^z + g \sigma_i^x ) - h_b(t) \sigma_1^z.
\end{equation}

This Hamiltonian falls in the class given by~\eqref{cftham}, as its low-energy excitations are described in equilibrium by the Ising conformal field theory. We note that the critical Ising model with a spatially disordered boundary field was studied in Ref.~\cite{Cardy_1991}.

After a Jordan-Wigner transformation~\cite{sachdev2007quantum}, the model (\ref{eq:ham})  maps onto a chain of free Majorana fermions 
\begin{equation}
    H(t)=-J\sum_{n=1}^{2L}i\eta_n\eta_{n+1}-h_b(t)\,i\gamma\eta_1,
    \label{eq:majorana}
\end{equation}
where $\eta_{2i-1}$, and $\eta_{2i}$ are Majorana operators located on site $i$ of the Ising chain. 
Note that expressing the boundary coupling to the edge operator $\sigma_1^z$, which breaks the Ising-symmetry, requires an additional ancilla Majorana operator $\gamma$ that anticommutes with all fields and satisfies $\gamma^2 = 1$~\cite{doi:10.1142/S0217751X94001552}. The quadratic Hamiltonian \eqref{eq:majorana} can be easily diagonalized numerically on large systems, {and is thus an ideal testbed for our earlier analytical arguments, which require large system sizes, late times and extensive disorder averaging to numerically observe.}

The system is endowed with three characteristic time scales:  the inverse bandwidth, $t_J\sim1/J$, which is the ultra-violet scale in the problem and controls the onset of non-universal effects in dynamics; the time-scale associated to the  boundary field $t_b = h_b^{-2}$;  and the intrinsic time of a stochastic Poisson flip, $\delta t$.  To ensure universal scaling, we choose  $t_J \ll t_b$, equivalent to the condition $h_b^2 \ll J$ (the boundary CFT limit). 
We note that if we were to integrate over the stochastic boundary field from the start, we would obtain an effective non-unitary evolution of a density matrix. However, because the Poisson switching process cannot be represented by a Gaussian white noise field, this not in general described by a quantum master equation in Lindblad form~\cite{PhysRevA.95.012115, Lucz}. Thus, the results presented here are distinct from previous works on driven-dissipative impurities, which used Lindblad equations to represent the drive~\cite{PhysRevLett.122.040604,PhysRevLett.122.040402, PhysRevB.85.184302, dries}. 

In our exact numerical calculations~\footnote{For details of the free-fermion numerical procedure, see the supplemental material of e.g. ~\cite{PhysRevLett.118.260602}, or the original references in~\cite{peschel_calculation_2003,eisler_evolution_2007}.},
we prepare the ground state of the chain and then  compute the time-dependent Loschmidt echo for at least 1000 realizations of the noise, on system sizes up to $L=1000$ and with $J=2$. At any given time step, we randomly select whether or not to flip the boundary field, corresponding to a Markovian process. We then scan over many values of the boundary field $h_b$ and the probability of flipping $p$ for two different types of drives: 1) a ``fixed-fixed'' drive, where the boundary field takes values $\pm h_b$ (with the system prepared in the ground state of $-h_b$), and 2) a ``free-fixed'' drive, where the boundary field takes values $+h_b$ and 0 (with the system in the ground state of $h_b=0$). Note that at very long-times we generally expect to see decay of the Loschmidt echo in any finite system as it heats up under the action of the incoherent drive, $h_b(t)$~\cite{Marinolong2012,cai}. However, this occurs on time scales of at least $t_{*}\propto L$~\cite{marko,PhysRevA.91.052107}, while in our simulations we keep $t < L/2$ to reduce finite-size effects, ensuring $t \lesssim t_*$.

Fig. ~\ref{fig:numerics} (left panel) shows the decay of the typical echo $\overline{\log\mathcal{L}}$ for different instances of the boundary field. The universal collapse, the asymptotic power law and the specific exponents obtained for both types of drive (fixed-fixed and free-fixed) are in excellent agreement with the CFT predictions. 

The right panel of Fig. ~\ref{fig:numerics} shows the results for the mean echo. As expected from the discussion of the previous section we see that the mean echo is identical to the typical echo in the case of the free-fixed drive. This is because the BCC operator has dimension $\Delta_{\mathrm{BCC}} = 1/16 < 1/4$ in this case. Again, as expected, the mean and typical echos differ substantially in the case of the fixed-fixed drive, for which the BCC operator $\Delta_{\mathrm{BCC}} = 1/2>1/4$. Furthermore, the inset shows reasonable data collapse with the ansatz 
$\overline{\mathcal L} \sim C_\lambda (T/t_b)^{-\lambda \alpha}$, where $\alpha=0.71\pm 0.03$ and $C_\lambda$ is a constant prefactor dependent on the Poisson rate of flipping. 
This should be compared to the analytical prediction of $\alpha=0.75$ obtained from our approximation above, taking into account only the leading divergences in the average over BCC insertions. Notice, however, that there is a larger statistical error in the average echo compared to the typical one; therefore, the imperfect collapse could either be due to  statistical errors or from actual small corrections to the  scaling exponent predicted from the bCFT analysis above.

\paragraph{Discussion.} The scaling exponents that control the dynamics of the Loschmidt echo in the critical transverse field Ising model are those of the boundary Ising CFT;  we therefore expect our results to hold upon adding integrability breaking perturbations $V$ to the Hamiltonian in~\eqref{eq:ham}, provided they are irrelevant operators under renormalization group flow (for instance, $V=\Gamma\sum_i\sigma^x_i\sigma^x_{i+m}$, with $m>0$).  
Furthermore, other critical points with central charge $c=1/2$ will give the same  dynamical scaling exponents. While we have demonstrated the scaling numerically for the Ising CFT, we emphasize that the mechanism for universality outlined here is model-independent. Any boundary-driven CFT  will display similar universal collapse when driven by appropriate boundary perturbations, with  exponents that depend on the particular form of the drive and driving operator. 
{We remark that the stochastic boundary Ising problem solved here does not map onto a Kondo problem (as done  in Ref.~\cite{PhysRevLett.100.165706}), since the average echo and the
mean echo studied in our work are not expressible as the statistical partition
function of a Coulomb gas.}

 An important general question is under what conditions one should expect to find universal behavior of a driven impurity. The problem of a quantum critical Ising chain driven by noise acting on a local transverse spin operator $h_x(t)\sigma^x_1$ was studied by one of us in Ref.~\cite{PhysRevLett.122.040604}. 
In that study, crucially, the critical Ising chain was driven by a \textit{marginal} boundary operator, $\sigma^x_1$, rather than by a \textit{relevant} boundary operator, $\sigma^z_1$.
 Despite this seemingly small difference, driving by a marginal spin operator yielded a decaying Loschdmit echo $\mathcal{L}(t)\propto e^{-\gamma t}t^\theta$, with a non-universal exponent $\theta$.
 This is in sharp contrast to the universal scaling collapse found in this work, and suggests that the  RG relevance of the driving operators can play an important role in dictating the universality (or lack thereof) of the dynamical response to dissipative impurities. 
Further, whether other classes of noise, such as $1/f$ noise or non-Markovian noise, can lead to novel dynamical universal scaling is an intriguing open question. Answering such questions would hopefully serve as stepping stones towards the goal of a systematic categorization of the universality classes of driven-dissipative impurities.

\begin{acknowledgments}
	\paragraph{Acknowledgements.} We acknowledge useful discussions with P. Fendley, M. Kolodrubetz, T. Prosen,  R. Vasseur and especially J. Cardy. WB is supported in part by the Hellman Foundation and by the DARPA
	DRINQS program (award D18AC00014). JM is supported by the European Union's Framework Programme for Research and Innovation Horizon 2020 under the Marie Sklodowska-Curie Grant Agreement No. 745608~(`QUAKE4PRELIMAT'). This collaboration was started during the KITP program `The Dynamics of Quantum Information', and it has been  supported in part by the National Science Foundation under Grant No. NSF PHY-1748958. 
\end{acknowledgments}

\bibliography{biblio}
\bibliographystyle{apsrev4-1}

\end{document}